\documentclass[fleqn,12pt,twoside]{article}
\usepackage{espcrc1}

\usepackage{graphicx}

\usepackage[figuresright]{rotating}

\newcommand{\AmS}{{\protect\the\textfont2
  A\kern-.1667em\lower.5ex\hbox{M}\kern-.125emS}}

\newcommand{\be}{\begin{equation}}    
\newcommand{\ee}{\end{equation}}    
    
\newcommand{\bef}{\begin{figure}}    
\newcommand{\eef}{\end{figure}}    
\newcommand{\bea}{\begin{eqnarray}}  
\newcommand{\eea}{\end{eqnarray}}

\def\spose#1{\hbox to 0pt{#1\hss}}    
\def\ltapprox{\mathrel{\spose{\lower 3pt\hbox{$\mathchar"218$}}    
 \raise 2.0pt\hbox{$\mathchar"13C$}}}    
\def\gtapprox{\mathrel{\spose{\lower 3pt\hbox{$\mathchar"218$}}    
 \raise 2.0pt\hbox{$\mathchar"13E$}}}    
\def\inapprox{\mathrel{\spose{\lower 3pt\hbox{$\mathchar"218$}}    
 \raise 2.0pt\hbox{$\mathchar"232$}}}

\newcommand{\bean}{\begin{eqnarray*}}
\newcommand{\eean}{\end{eqnarray*}}
 
\title{Statistical Physics for Cosmic Structures}

\author{Luciano Pietronero \address[ROMA]{INFM Sezione Roma1, 
        Dip. di Fisica, Universit\'a ``La Sapienza'', \\ 
        P.le A. Moro, 2, I-00185 Roma, Italy.},
        Andrea Gabrielli \addressmark,
        and
       Francesco Sylos Labini\address[PARIS]{Laboratoire de 
Physique Theorique, Unversit\'e de Paris XI, 91405 Orsey Cedex France}
       \addressmark[ROMA]}

\begin{document}

\maketitle

\begin{abstract}
Ideas of Statistical Physics are very relevant for cosmic structures 
especially considering that the field is undergoing a period of exceptional 
development with many new data appearing on a monthly basis.
In the past years we have focused mostly on galaxy distributions and their 
statistical  properties.
This led to an interesting debate which will be resolved by the next
generation of data in a couple of years.
In addition here we discuss the statistical properties of the fluctuations
of the cosmic microwave background which are small in amplitude but complex in
structure.
We finally discuss the connection between these observations and the 
Harrison-Zeldovich spectrum and its further implications on the theories of 
structure formation and the and the cosmological $N$-body simulations.
\end{abstract}

\section{Introduction}

From the point of view of Statistical Physics cosmic structures give rise to 
a puzzling and apparently contraddictory situation:
on one hand the Cosmic Microwave Background Radiation 
(hereafter CMBR) is extremely
isotropic with temperature fluctuations of the order 
$\delta T/T \simeq 10^{-5}$ \cite{cobe,boom}.
On the other hand matter distribution is extremely clumpy: there are galaxies,
clusters, filaments and large voids up to huge scales \cite{cp92,slmp98}.
Fluctuations with a large density contrast $\delta \rho/\rho\gg 1$
appear to characterize the matter field over a broad range of scales.
The link between the two observations is, of course, not direct.
The first one refers to radiation which is essentially unchanged from
the time of the electro-magnetic decoupling.
This is when the universe was about $1000$ times younger than now.
Galaxies instead correspond to the present (visible) matter field.
Another important difference is that the CMBR corresponds to an angular
distribution, while the galaxy distribution is three-dimensional, in view
of the red-shift ($z$) measurement.
Before the decoupling time, matter and radiation are coupled by the 
Sachs-Wolf effect \cite{pee93} 
which describes the fact that if radiation escapes from a 
gravitational potential well it is red-shifted by an amount linear
with the potential.

This implies that, at $z=1000$ (Universe $1000$ times younger than now) 
$\delta \rho/\rho\simeq \delta T/T\simeq 10^{-5}$.
The question is then if gravitational instability can lead to the present
large structures, starting from $\delta \rho/\rho\simeq 10^{-5}$
at $z=1000$.
The answer is no, in fact one could get basically an amplification
of a  factor $10^3$
by gravity, in the linear regime, 
leading to a present value of the matter fluctuations
$\delta \rho/\rho\simeq 10^{-2}$, in sharp contrast with the large 
structures observed.
This is one of the main reason to introduce a  vast amount of 
non-baryonic dark matter 
matter (it interacts very weakly with radiation) which is supposed to have,
at the decoupling time, fluctuations much larger than $10^{-5}$.
In view of this and other puzzling observations the current interpretation
is that the matter field consists in about $99\%$ of the total matter
and the baryonic matter is just about $1\%$.

From this discussion it is clear that the understanding of the nature 
of the various cosmic structures and their dynamical development
is a central point of the field which has implications in essentially all 
other properties.
Many new data have been generated in the past few years and many more are 
expected in the near future.
So the field is quickly evolving from a situation with many speculations 
and few data to a new one, in which the new abundance of data permits a test
of long standing conjectures.
This is therefore an extremely interesting moment in which ideas
from Statistical Physics can play a crucial role.
This is true grasping the properties of structures from the data 
but also in the theories  of structures formation and in the $N$-body
simulations.

In the past we have mostly considered the properties of the galaxy 
distributions \cite{cp92,slmp98}. 
Here we start instead from an analysis of the fluctuations corresponding 
to the microwave background radiation, and then we try to connect them
to the galaxy properties.
We outline the main open problems from a new perspective and we consider
some consequences also on the theories and 
cosmological $N$-body simulations \cite{bottaccio01,thierry}.

\section{Small fluctuations and Poisson distribution}  

The standard approach in the field is to assume that the matter field
$\rho (\vec{r})$ is always characterized by a well defined average density 
with small amplitude fluctuations.
In our opinion this may be acceptable in relation to the microwave radiation,
but not for the galaxy distribution.
We will come back on this point later.

The basic assumption is to have
\begin{equation}
\rho (\vec{r})=\rho_0 + \delta\rho (\vec{r})\;\;\mbox{with}\;\;
\delta\rho (\vec{r})\ll \rho_0\,,
\label{2.1}
\end{equation}
where $\rho_0$ is the average density.
This is then analyzed in terms of the correlation function of the 
fluctuations (positive and negative)
\be
\xi (r)=\frac{\left<\rho (\vec{r}_0)\rho (\vec{r}_0+\vec{r})\right>}
{\rho_0^2}-1 \equiv \frac{
\left<\delta\rho (\vec{r}_0)\delta\rho (\vec{r}_0+\vec{r})\right>}
{\rho_0^2}\,,
\label{2.2}
\ee
the normalized mass variance in spheres of radius $R$
\be
\sigma^2(R)=\frac{\left<M^2(R)\right>}{\left<M(R)\right>^2}-1,
\label{2.3}
\ee
(where $M(R)$ is the mass in a given sphere of radius $R$) and
the power spectrum of the normalized density fluctuations
which are supposed to have small amplitude with respect
to the average density
\be
P(k)=\left<|\hat{\delta} \rho(\vec{k})|^2\right>=
\frac{1}{(2\pi)^3}\int d^3r\,e^{-i\vec{k}\vec{r}}\xi(r)\,,
\label{2.4}
\ee
where $\hat{\delta} \rho(\vec{k})$ is the Fourier transform
of the normalized density contrast field $\delta\rho(\vec{r})/\rho_0$.

It is very instructive to consider various properties and in particular
the fluctuations of the gravitational potential corresponding to a
Poisson distribution of mass points.
The Poisson distribution is completely random and uncorrelated.
In a sphere of radius $R$ we have an average number of points given by
\be
\left<N(R)\right>=\rho_0 V\sim \rho_0 R^3\,,
\label{2.5}
\ee
where $V\sim R^3$ is the volume of the sphere in three dimensions.
The typical number fluctuation in such a sphere is given by
\be
\delta N(R)=\left<N(R)\right>^{\frac{1}{2}}\sim R^{\frac{3}{2}}\,.
\label{2.6}
\ee
With respect to our functions we have that in the thermodynamic limit
of an infinite distribution of points
\be
\xi(r)=\delta(\vec{r})\;\;;\;\;P(k)\sim k^0\sim const. >0\,.
\label{2.7}
\ee
However in a finite sample, for the finite size fluctuations due to
Eq.(\ref{2.6}), for a generic statistical estimator of $\xi(r)$
we have
\be
|\xi(r)| \simeq \left|\frac{\tilde{\delta\rho}}{\rho_0}\right|\simeq
\left|\frac{\delta N}{\left<N\right>}\right|\simeq 
\left<N\right>^{-\frac{1}{2}}\sim r^{-\frac{3}{2}}
\label{2.8}
\ee
for the absolute value of $\xi(r)$, where $\tilde{\delta\rho}$ is the 
typical fluctuation of the conditional density from an occupied point
\footnote{Note that there are some special statistical estimators of
$\xi(r)$, for the particular case of a Poissonian distribution only, giving
$|\xi(r)|\sim r^{-3}$.}.
For $P(k)$ the finite value of $N$ induces some noise on top of
the constant value of Eq.(\ref{2.7}).

Now comes the key-point because we can estimate the fluctuations of the 
gravitational potential $\delta \phi (R)$ as a function of the scale
$R$:
\be
\delta\phi(R)\sim \frac{\delta N(R)}{R}  \sim
R^{\frac{1}{2}}\,.
\label{2.9}
\ee
So we have diverging fluctuations of the potential originated by the finite 
size fluctuations of the Poisson distribution Eq.(\ref{2.6}).
This is a very important point which has deep implications
 with respect to the 
Harrison-Zeldovich spectrum (see below).
However, if we estimate in the same way the fluctuations of the gravitational
force $\vec{F}(R)$ we have the surprising result that they are instead bounded
\be
|\vec{F}(R)|\sim |\nabla\delta\phi(R)|\sim R^{-\frac{1}{2}}\,.
\label{2.10}
\ee
This result, that the gravitational force is well defined in a 
Poisson distribution of points, is well known since the seminal work of 
Chandrasekhar \cite{chandra}. The extension of this result to the
fractal
case is rather complex, and the interested reader can find a discussion
in \cite{gslp99}.

These properties of the gravitational field of a random distribution of
of massive points are usually not very appreciated but they already
pose, in our opinion, extremely interesting and challenging problems 
as we are going to see in the following.

\section{The Harrison-Zeldovich (HZ) spectrum} 

\begin{figure}[htb]
\includegraphics[width=40pc,height=20pc]{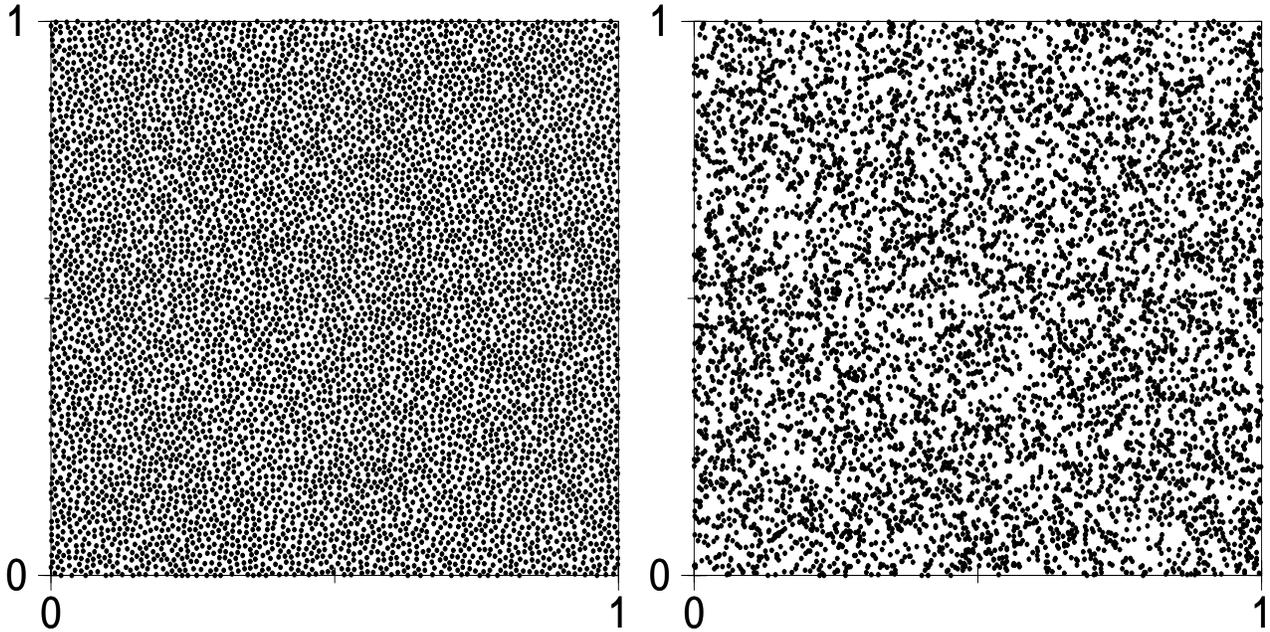}
\caption{
{\em Left panel}:  Glass-like distribution of points
obtained by the method of the shuffled lattice \cite{hz}.
The idea is to decrease the fluctuations below to the level of the 
Poisson random distribution. This achieved by imposing a maximal value  for
the nearest neighbors distance. In this case the fluctuation is about
$\delta N(R) \sim \left<N(R)\right>^{1/3}$ and it is a qualitatively
good representation of the HZ spectrum which in this respect can be 
considered {\em super-homogeneous} because it is more homogeneous 
than a Poisson distribution. 
{\em Right panel}: Example of a Poisson distribution. 
In this case the behavior of the typical fluctuation of the number of points
in a ball of radius $R$ goes like $\delta N(R) \sim \left<N(R)\right>^{1/2}$.
The distribution is homogeneous at large scale with a well defined average 
density.}
\label{fig1}
\end{figure}

\begin{figure}[htb]
\includegraphics[width=40pc,height=20pc]{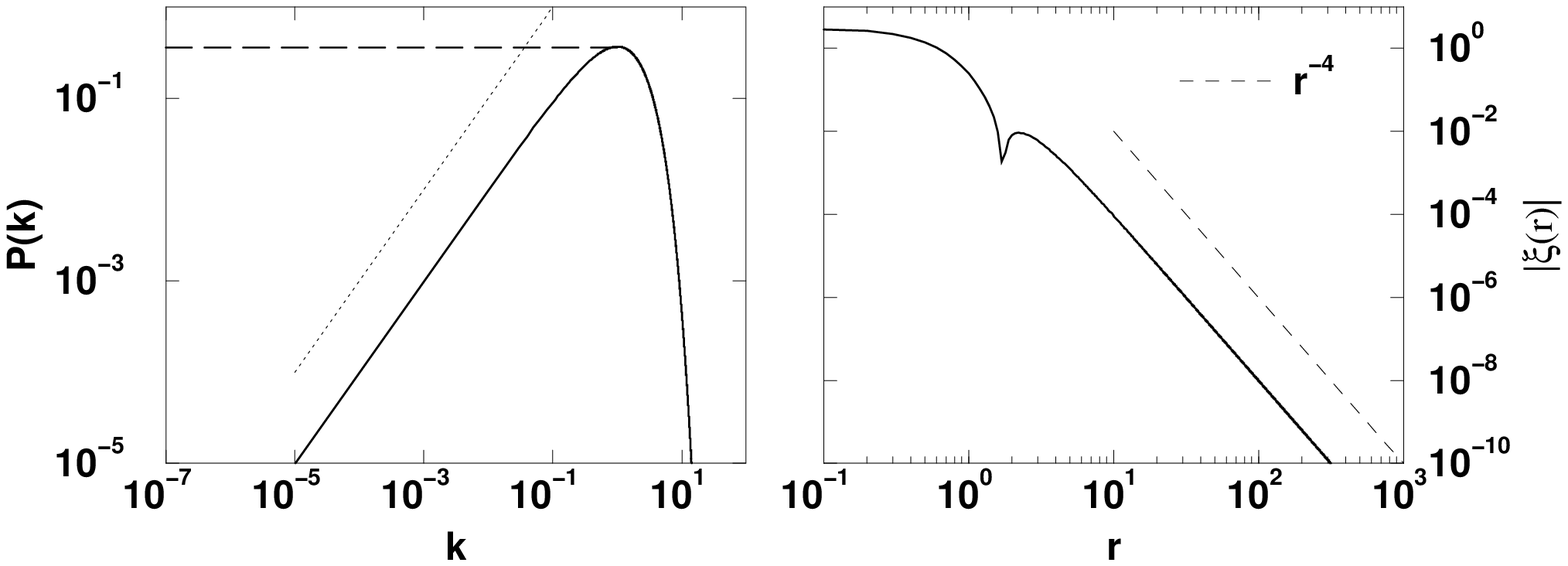}
\caption{
{\em Left panel}: Power spectrum corresponding to the HZ spectrum
(solid line).
Note that the behavior at large scale (small $k$) $P(k)\sim k$
(dotted line)
implies less fluctuations than the Poisson behavior $P(k)\sim
k^0\sim const.>0$ (dashed line).
{\em Right panel}: Modulus of the correlation function $\xi(r)$
(solid line)
corresponding to the previous power spectrum. Note that at large scales
the function is negative (the reference dotted line 
has a slope $r^{-4}$). Note that the units are arbitrary.}
\label{fig2}
\end{figure}

We consider now the standard derivation of the HZ spectrum of primeval
fluctuations which has a central role in the field.
The idea is to consider the implications of the CMBR observations as given
for example by the COBE experiment \cite{cobe}.
In this experiment the temperature fluctuations of the black body spectrum 
are measured at large angular separation ranging from $7$ to $90$ degrees.
The observation is that the value $\delta T/T\simeq 10^{-5} $
is about the same for small and large angles.
This means that this fluctuation does not grow with the angle.
The standard conjecture (i.e. \cite{pee93}) 
is that one can go from angles 
to distances and argue that
\be
\delta T(\theta)\sim \delta \phi (R)\sim const\,.
\label{3.1}
\ee
Namely, if the temperature fluctuation is angle independent, then
the gravitational potential fluctuation which, via the Sachs-Wolf effect
led to the temperature fluctuation, should be independent on the scale.
This is sometimes defined as a conditional ``scale invariance'', but
instead is just that the potential fluctuations are simply constant at 
various scales \cite{hz}.
We can immediately see that a Poisson distribution does not satisfy
this condition because the fluctuations of the gravitational potential
grow with scale as in Eq.(\ref{2.9}).
It is easy to see that, in order to satisfy the Eq.(\ref{3.1}), one
should have
\be
\delta N(R) \sim \left<N(R)\right>^{\frac{1}{3}}\,,
\label{3.2}
\ee
which implies a more regular distribution ({\em super-homogeneous}) with
respect to a random Poisson one (see Fig.~\ref{fig1}).
With respect to the correlation function and the power spectrum, it can be 
easily shown that at large scales \cite{hz} 
\bea
\label{3.3}
\xi(r)&\sim& -r^{-4}\\
\label{3.4}
P(k)&\sim& k\,.
\eea
Note the negative sign for the large scale behavior of $\xi(r)$:
this means anti-correlations at large scales (see Fig.~\ref{fig2}).

This discussion leads to a series of questions about the implications 
of the density fluctuations on the gravitational potential which, in 
our opinion, would require some deeper clarifications:
(i) a Poisson distribution has a perfectly well defined average density
at large scales $\rho_0=\lim_{V\rightarrow\infty} N(V)/V$.
Consequently, one could suppose that at large scale one
can use the value $\rho_0$ in general relativity and describe it
in terms of the Friedman metrics.
However, given our discussion and Eq.(\ref{2.9}), this should not be 
permitted because the finite size fluctuations $\delta N\sim
\left<N\right>^{1/2}$ induce  diverging gravitational potential
fluctuations even at large scales where $\delta N\ll \left<N\right>$.
If, on the other hand, one would compare forces and not potentials, 
then the perturbation would be negligible.
So the question to clarify is whether a Poisson distribution of mass 
points is compatible with a Friedman cosmology at large scales.
(ii) The same reasoning holds for the derivation of the HZ spectrum
as necessarily implied by the observed properties of the CMBR.
Note that the HZ spectrum is so central to the field that this and
the previous points should either be reinforced on disproven.
(iii) Other points concern the relations between angular and space 
correlations.
We have learned that for galaxies the two properties can be very different
due to the projection effects of the angular case.
Finally the Sachs-Wolf effect for non trivial density fluctuations should
be considered.

\section{Galaxy correlations updated}

For galaxy correlations all the above concepts of small fluctuations 
theory etc. are certainly not appropriated.
One observes fractal clustering with large scale structures and voids
for which $\delta \rho/\rho_0\gg 1$ or even worse $\rho_0$ is not
well defined in the available samples.
This is the field in which we have worked extensively and proposed a radically
new and more general way to interprete the puzzling observations and
to define the theoretical perspective.
One of the consequence of our work is that the so called characteristic
length of galaxies $r_0\simeq 5 Mpc/h$ \cite{pee93} 
is an artefact of the {\em a-priori}
assumption of homogeneity and, in our opinion, correlations extend
up to the present limits of observations \cite{cp92,slmp98,jsl01}.
In any case there is, finally, agreement on the fact that at least to some
small scale, galaxy correlations are fractal.
A common misunderstanding is however to think that if there would be a 
crossover, then the small fluctuations theory could describe the fractal 
properties because they extend only on a finite range of
 scale (e.g. \cite{rees98}).
This is clearly incorrect because, independently of the presence of the
crossover, a theory based on small fluctuations with 
$\delta \rho/\rho_0\ll 1$
can never be appropriated to describe fractal clustering.

Apart from this conceptual point, which is anyhow valid, the question
of the extension of the fractal properties should be clarified by next
catalogues. In the next few years the new catalogs like 2dF and SDSS
will dramaticaly improve our knowledge of the large scale
distribution of galaxies. The ongoing debate in the field 
\cite{dav97,pmsl97,rees98,jsl01} will be then clarified 
by the analysis of these new data.

It is clear therefore that, at least for the properties of galaxy clustering,
the formalism of the small fluctuation theory is not appropriate at any stage
from the data analysis to the formulation of a theory of structure formation.

\section{Conclusion and perspectives}

We have presented a brief overview of 
the relevance of Statistical Physics
for cosmic structures.
Many new data are appearing for the CMBR and for the galaxy distribution, and
modern Statistical Physics is the natural framework to identify the relevant 
structures  and defining the corresponding theoretical models.

In this respect the situation appears quite different for the very small 
fluctuations  in the CMBR and the very large fractal fluctuations observed
for galaxies. In the first case the traditional theory of small fluctuations 
could be appropriated, but it is certainly not appropriated for galaxies.
Whatever clustering is observed this has scaling and fractal properties 
and, even if there would be a crossover to an homogeneous distribution
(a fact which is still far from shown) the fractal clustering
could never be properly described by a small amplitude fluctuations theory.

A different question concerns the cosmological model able to 
describe a situation where matter has fractal properties.
In Ref.\cite{pwa} the interested reader can find an attempt 
to construct a Friedman model with zero density,
where the fractal can be treated as perturbation to the 
background homogeneous metric.  

It is therefore unavoidable to generalize the approach to include the 
possibility of complex structures and large fluctuations at alla levels,
from the data analysis to the theoretical methods and the $N$-body
simulations \cite{bottaccio01,thierry}.
This should lead to a new general perspective of the field, strongly based
on a cross-disciplinary approach, which certainly will lead to novel
exciting developments.

\end{document}